\documentclass[aps,pra,twocolumn,amssymb,amsmath,superscriptaddress,reprint]{revtex4-1}
\usepackage[colorlinks=true,urlcolor=blue,citecolor=blue,linkcolor=blue]{hyperref}
\usepackage{graphicx,amsmath,amssymb,amscd,epsfig}
\usepackage{color}
\usepackage{placeins}
\usepackage{float}

\begin{document}

\author{Nikita~V.~Tepliakov}
\email{nikita.tepliakov@epfl.ch}
\affiliation{Institute of Physics, Ecole Polytechnique F\'{e}d\'{e}rale de Lausanne (EPFL), CH-1015 Lausanne, Switzerland}
\affiliation{Information Optical Technologies Center, ITMO University, Saint Petersburg 197101, Russia}
\author{QuanSheng~Wu}
\author{Oleg~V.~Yazyev}
\email{oleg.yazyev@epfl.ch}
\affiliation{Institute of Physics, Ecole Polytechnique F\'{e}d\'{e}rale de Lausanne (EPFL), CH-1015 Lausanne, Switzerland}

\title{Crystal Field Effect and Electric Field Screening in Multilayer Graphene \\ with and without Twist}

\date{\today}

\begin{abstract}
We address the intrinsic polarisation and screening of external electric field in a broad range of ordered and twisted configurations of multilayer graphene, using an \textit{ab initio} approach combining density functional theory and the Wannier function formalism. We show that multilayer graphene is intrinsically polarized due to the crystal field effect, an effect that is often neglected in tight-binding models of twisted bilayer graphene and similar systems. This intrinsic polarization of the order of up to few tens of meVs has different out-of-plane alignments in ordered and twisted graphene multilayers, while the in-plane potential modulation is found to be much stronger in twisted systems.
We further investigate the dielectric permittivity $\varepsilon$ in same multilayer graphene configurations at different electric field strengths. 
Our findings establish a deep insight into intrinsic and extrinsic polarization in graphene multilayers and provide parameters necessary for building accurate models of these systems. 
\end{abstract}

\maketitle

Graphene and its derivatives continue attracting attention due to their ever-growing list of novel physical properties and potential technological applications \cite{Novoselov04,CastroNeto09,Novoselov12}.
Multilayer graphene (MLG) is particularly eminent owing to the exceptional versatility of its design -- it is possible to vary the number, stacking order or twist between graphene layers to tailor electronic and optical properties \cite{min2008electronic,koshino2008magneto,tepliakov2020twisted}. In particular, electronic properties of twisted multilayer graphene (tMLG) are strongly affected by the moir\'{e} superlattice potential resulting from the relative rotation of the layers.
\cite{Laissardiere2010, morell2010flat, Shallcross2010, Bistritzer2010,mele2010commensuration}
Notably, bilayer graphene twisted by {\it ca.} $1^\circ$, the so-called magic angle, features flat bands at the charge neutrality point resulting in the recently discovered superconducting and correlated insulator phases \cite{cao2018unconventional,cao2018correlated}. 
Similar physics has also been revealed in more complex configurations, such as twisted graphene trilayers \cite{Chen19} and twisted double bilayer graphene \cite{Shen20,Cao20,Liu20}. This renders tMLG a promising platform for exploring correlated electron behavior with unprecedented tunability and precision. 
Electronic properties of MLG can further be controlled by means of external electric potentials.
This provides a simple avenue for controlling the charge-carrier concentration by gating. In addition,
the out-of-plane electric field 
has a strong effect on the band structure \cite{williams2012tunable}, \textit{e.g.} opening a gap in the electronic spectrum of the Bernal-stacked bilayer graphene \cite{Zhang09,Oostinga08,castro2007biased,mccann2006asymmetry}. 

Exploring novel physical phenomena and designing graphene-based devices relies on modelling tools that combine accuracy of predictions with computational efficiency. The tight-binding approximation that considers only the $p_z$ orbitals of carbon atoms combined with the Slater-Koster formalism for describing interlayer hopping integrals \cite{Lisardiere10} is the most commonly used model allowing to treat systems of sufficiently large size, {\it e.g.} the magic-angle twisted bilayer graphene.  
While this approximation works well in many cases, it usually does not account properly for the crystal fields which appear in MLG due to the different atomic environments of individual atoms. 
For example, 
accounting for different environments in the tight-binding description has proved crucial for the correct description of twisted double bilayer graphene \cite{rickhaus2019gap,haddadi2020moire,Culchac20}.
Therefore, accurate description of various MLG configurations requires the knowledge of crystal field parameters. 

Furthermore, the modelling of biased MLG-based devices requires an understanding of the electric field screening in this material \cite{sun2010spectroscopic}. As one of the first steps in this direction, Koshino has built a self-consistent tight-binding model of field penetration in MLG \cite{koshino2010interlayer}. 
Most importantly, this model revealed qualitatively different dielectric response of Bernal and rhombohedral-stacked graphene multilayers. A similar model was developed for doped MLG, where the layers were assumed to interact only electrostatically via excess charges \cite{kuroda2011nonlinear}, while a first-principles study of MLG revealed that its dielectric permittivity grows with the electric field strength due to charge polarization \cite{santos2013electric}. These works, however, do not provide a complete picture covering the twisted MLG configurations.

In this Letter, we address both aforementioned problems by studying the crystal-field effect and the screening of perpendicular electric field in different MLG configurations. First, we perform first-principles calculations of both \textit{ordered} and \textit{twisted} MLG configurations composed of up to 8 layers, with and without an explicitly applied external electric field of varying strength. We further construct the projector Wannier functions using the $p_z$ atomic orbitals of MLG. This procedure yields \textit{ab initio} tight-binding Hamiltonians, from which the crystal field parameters and dielectric constants are obtained. 
We show that MLG configurations consisting of ordered ABA and ABC stacking sequences feature an intrinsic symmetric polarization field, which has opposite direction to that in the twisted systems. We also find that the crystal field within the graphene layers in tMLG systems is strongly modulated by the moir\'{e} potential. We then calculate the dielectric permittivity of ordered and twisted MLG using the same approach. The dielectric permittivity of the ordered MLG is found to be strongly dependent on the electric field strength, ranging between $2.3$ and $5.4$ in the case of Bernal-stacked multilayers and between $2.0$ and $4.3$ in the case of rhombohedral stacking. In contrast, tMLG has a practically constant dielectric response characterized by larger permittivity values varying between $3.4$ and $5.8$ depending on the number of layers. 
The crystal field parameters and dielectric permittivities obtained in this work can be readily introduced into the tight-binding and other semiempirical models to ensure accurate description of the physical phenomena in various MLG configurations.

\begin{figure}
	\centering
	\includegraphics[width=0.45\textwidth]{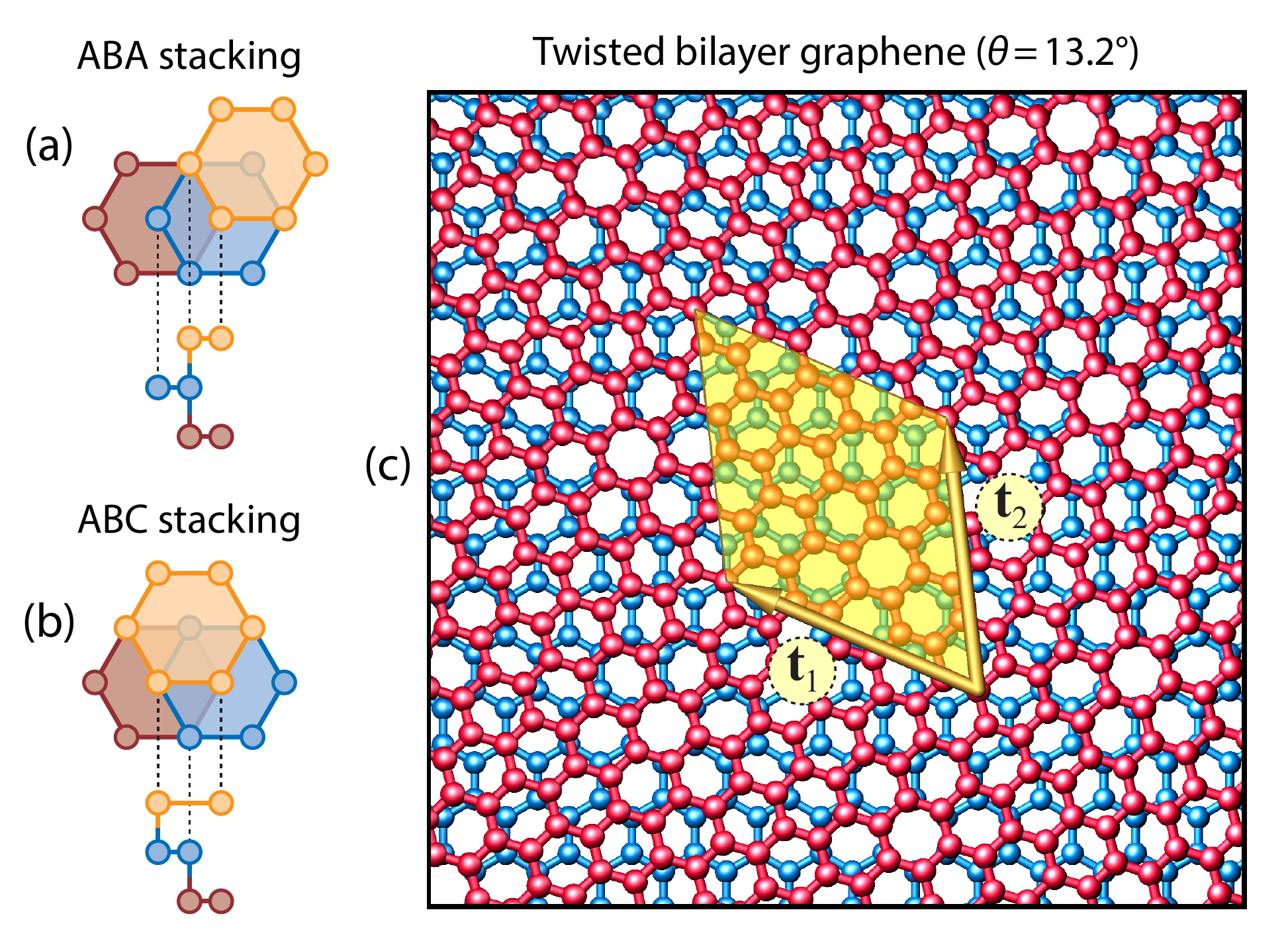}
	\caption{Two-dimensional crystal lattices of (a) ABA-stacked trilayer graphene, (b) ABC-stacked trilayer graphene, and (c) TBG with twist angle $\theta\approx13.2^\circ$. Panel (c) shows the moir\'{e} unit cell defined lattice vectors $\mathbf{t}_1$ and $\mathbf{t}_2$.}
	\label{geometry}
\end{figure}

In this study, we focus on two different types of MLG configurations. The first type represents \textit{ordered} configurations of up to $N=8$ layers arranged in Bernal (ABA) and rhombohedral (ABC) stacking sequences, which are shown in Figure~\ref{geometry}(a) and (b). 
The Bravais lattice vectors of the ordered system are given by $\mathbf a_1 = (a,0,0)$ and $\mathbf a_2 = (-a/2,\sqrt{3}a/2,0)$, where $a=2.46$~\AA\ is the lattice constant of graphene. The distance between layers $d=3.35$~\AA\ is assumed to be equal to that in bulk graphite. 

Upon rotating graphene layers with respect to each other, the periodicity in the $xy$-plane is preserved only for a specific set of commensurate twist angles given by
\begin{equation}\label{angle}
	\cos\theta = \frac{3i^2+3i+1/2}{3i^2+3i+1},
\end{equation}
where $i=1,2,3,\dots$. The lattice vectors of such \textit{twisted} multilayer graphene are given by $\mathbf t_1 = i\mathbf a_1 + (2i+1)\mathbf a_2$ and $\mathbf t_2 = -(i+1)\mathbf a_1 + i\mathbf a_2$, as illustrated in Figure~\ref{geometry}(c). In this work, we study only the twisted structures with an even number of layers $N=2,4,6,8$ and restrict our consideration only to twists in the middle of the multilayer, so that the top half of the layers is rotated with respect to the bottom half.
The stacking order in the two counterparts is of Bernal type. 

\textbf{Crystal-Field Effect.} Even in the absence of external electric field, multilayer graphene features an intrinsic polarization, which appears due to the different atomic environments of each lattice site. In order to estimate this intrinsic polarization from first principles, we employ an original approach combining density functional theory (DFT) calculations with the Wannier function formalism (see section Methods for details as well as Refs.~\cite{scaramucci2015crystal,pasquier2019crystal}). In this approach, the projector Wannier functions for the $p_z$-orbitals of carbon atoms are constructed from DFT wavefunctions to obtain an \textit{ab initio} tight-binding Hamiltonians of the considered MLG models. The diagonal elements of these Hamiltonian matrices correspond to the onsite potentials of the $p_z$-orbitals of carbon atoms. We denote those as $V_{nm}$, where $n$ is the layer number and $m$ is the atomic number within the layer. These onsite potentials are generally different as a manifestation of the crystal field effect.

It is instructive to decompose the crystal field effect into the out-of-plane and in-plane components, taking advantage of the planar geometry of multilayer graphene. The out-of-plane crystal field, which we shall refer to as the intrinsic symmetric polarization (ISP), is calculated by averaging the on-site potentials within each layer, $V^\mathrm{out}_n=\sum_m V_{nm}/N_\mathrm{pl}$, where $N_\mathrm{pl}$ is the number of atoms per layer. The resulting on-layer potentials are generally different and lead to intrinsic electric field perpendicular to graphene layers. Note that the symmetry of ISP implies that it can only be observed in multilayer graphene with $N\geq3$ layers, since the on-layer potentials in bilayer graphene are the same. The in-plane component of the crystal field is then found by subtracting the on-layer potentials from the atomic energies, $V^\mathrm{in}_{nm} = V_{nm} - V^\mathrm{out}_n$, which allows to characterize the crystal-field splitting within each layer independently from the rest of the structure.

\begin{figure*}[t]
	\centering
	\includegraphics[width=0.85\textwidth]{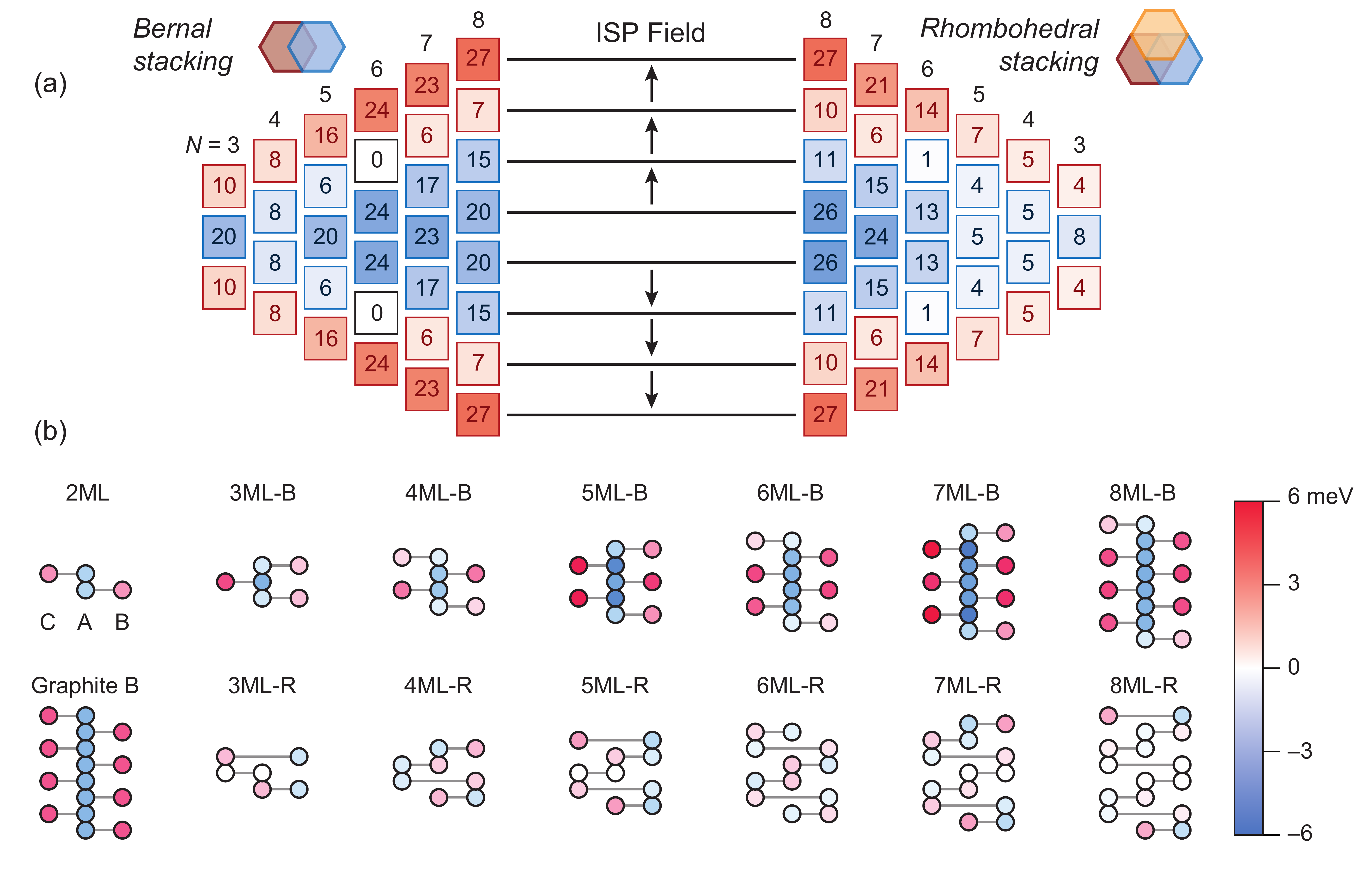}
	\caption{(a) Intrinsic symmetric polarization (out-of-plane crystal-field splitting, in meV) in graphene multilayers of two different stacking configurations and (b) in-plane crystal-field splitting parameters for multilayer graphene and the Bernal-stacked bulk graphite. In panel (a), red and blue correspond to the positive and negative values of the on-layer potentials, and the color intensity illustrates their magnitude (in meV). In the case of in-plane splitting in graphene, the panels show the lateral view of atoms in the unit cell.}
	\label{crystalOrdered}
\end{figure*}

The ISP in ordered multilayer graphene is summarized in Figure~\ref{crystalOrdered}(a). One can see that regardless of the number of layers or the stacking pattern, the ISP field is always directed towards the exterior layers of the structure. The maximum ISP values are found to be 30~meV of interlayer splitting in case of Bernal-stacked trilayer graphene. Overall, the ISP values for ABA-stacking configurations tend to be larger than those for the ABC stacking, due to the more pronounced asymmetry in the packing of atoms in Bernal graphene. The ISP values for both types of stacking are reproduced with greater precision in Tables~S1 and S2 in the Supplementary Information. 

In the Supplementary Information, we also use a simple self-consistent model of trilayer graphene to show that the ISP field can be associated to the hopping of electron between the exterior layers of the structure. This hopping redistributes electron density in the $z$ direction, leaving a positive charge vacancy inside the multilayer. By attracting electrons towards this vacancy, ISP field serves as a self-consistent correction accounting for such redistribution of the electron density.

\begin{figure*}[t!]
	\centering
	\includegraphics[width=0.75\textwidth]{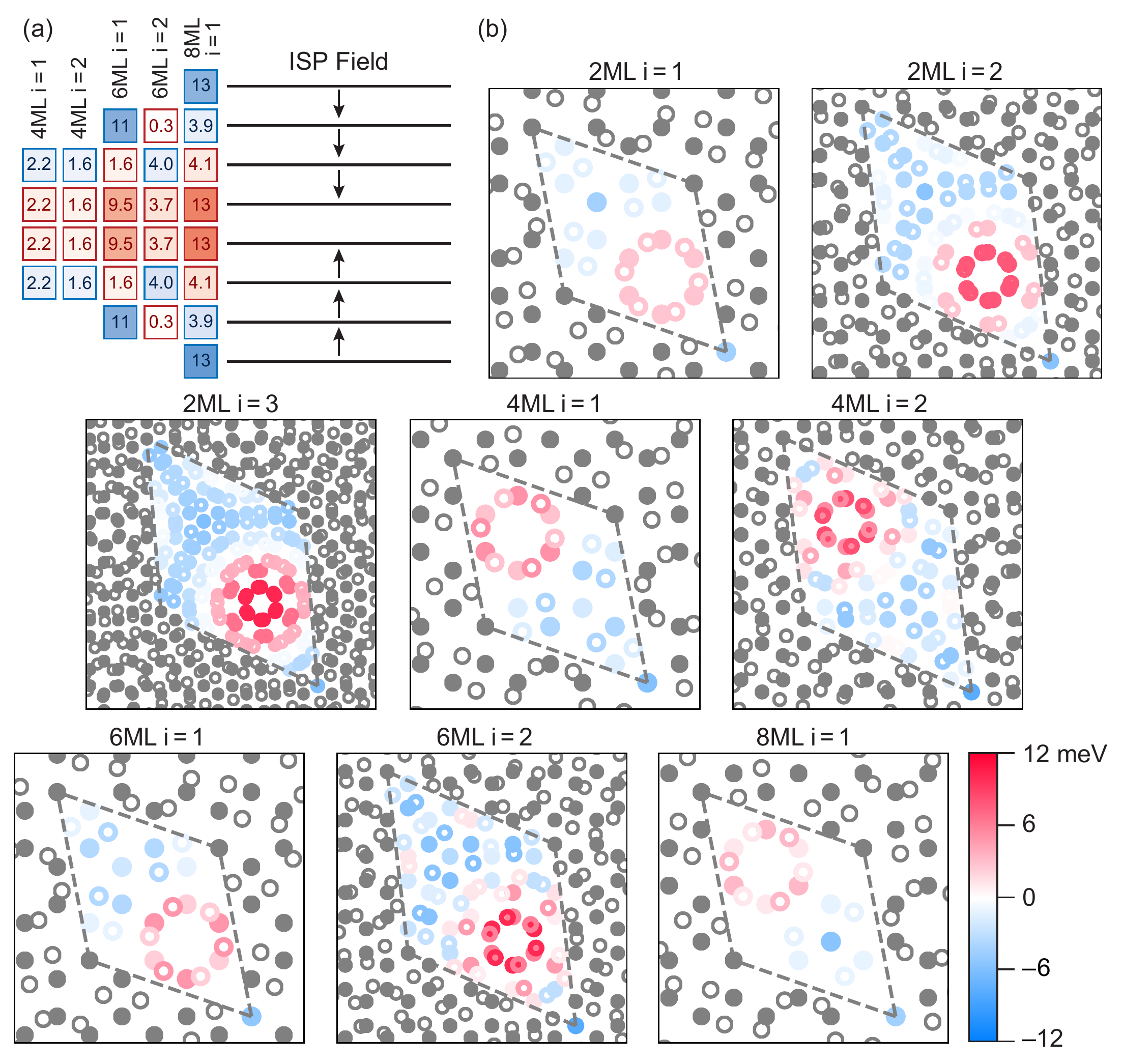}
	\caption{(a) Intrinsic symmetric polarization (out-of-plane crystal-field splitting, in meV) and (b) in-plane crystal-field splitting parameters in twisted multilayer graphene of 2, 4, 6 and 8 layers. Indices $i=1$, $2$, and $3$ correspond to twist angles $\theta=21.8^\circ$, $13.2^\circ$ and $9.4^\circ$. For all studied structures, the in-plane crystal field effect is shown at the twist boundary, \textit{i.e.} for the middle two layers. Open and filled circles distinguish the two graphene layers.}
	\label{crystalTwisted}
\end{figure*}

The in-plane crystal-field splitting parameters for the ordered graphene multilayers and bulk Bernal-stacked graphite are shown in Figure~\ref{crystalOrdered}(b). As follows from this figure (and Tables S3 and S4 in the Supplementary Information), the in-plane crystal-field effect in multilayer graphene is significantly weaker than the out-of-plane polarization, having typical values around 5~meV. Like in the case of the out-of-plane crystal field, ABA multilayers are characterized by a more pronounced in-plane polarization than the ABC-stacked MLG configurations.
As the number of layers increases, the in-plane splitting inside the ABA-stacked MLG naturally starts resembling the corresponding picture in the Bernal-stacked bulk graphite. On the contrary, the in-plane crystal-field effect in the ABC-stacked configurations is practically negligible for thick multilayers. The latter observation is consistent with the fact that in bulk rhombohedral graphite all atoms are equivalent and there is no crystal-field effect. Consequently, the two types of structures presented here are characterized by the qualitatively different in-plane crystal-field effects. While in the ABA-stacked multilayers the splitting is stronger inside the structure, in the ABC configurations it is only present in the surface layers.

The out-of-plane crystal-field effect for the twisted multilayers is illustrated in Figure~\ref{crystalTwisted}(a) and reproduced in Table S5 in the Supplementary Information. One can immediately observe that the ISP field direction in twisted systems is generally opposite to that in the ordered multilayers. 
According to our effective model of the ISP in graphene, this inversion of the intrinsic field is due to the different sign of the hopping between separated layers, as opposed to ordered graphene.
The values of the out-of-plane polarization are also substantially lower in twisted multilayer graphene compared to the ordered structures. The largest value, observed for the six-layer MLG with  $\theta=21.8^\circ$, is 12.6~meV of the interlayer splitting.
Such small values in twisted multilayers can be explained by the redistribution of the electron density within the moir\'{e} pattern, which is a more pronounced effect than the redistribution of electrons along the $z$-direction. It is also worth noting that the direction of ISP partially reverses in the six-layer layer graphene with twist angle $\theta=13.2^\circ$. It may be expected that in tMLG configurations with larger moir\'{e} supercells, the ISP field would recover direction observed for the ordered structures. 

Figure~\ref{crystalTwisted}(b) shows that the in-plane crystal-field splitting in tMLG is much more pronounced than in ordered MLG configurations and is comparable to the ISP fields in the twisted systems. For twisted bilayer graphene with twist angle $\theta=9.4^\circ$, the in-plane splitting at the twist boundary reaches 12~meV, which is twice as large compared to Bernal-stacked ordered configurations. In general, this in-plane crystal field increases as the moir\'{e} superlattice constant increases, while the number of layers has little effect on it.

In case of TBG [the first three panels of Figure~\ref{crystalTwisted}(b)], one can clearly identify a pattern in the in-plane crystal-field effect. The positive onsite potentials are centered on the regions of moir\'{e} lattice where the local stacking resembles that of the AA bilayer graphene, whereas the negative potentials are centered at the AB and BA stacking regions. In the Hartree formalism, this in-plane crystal field is attributed to the localization of electrons near the Fermi level in the AA-stacked regions of TBG \cite{goodwin2020hartree}. The localization enhances electron--electron interactions and increases the Hartree potential energy in the AA-stacked regions of the moir\'{e} crystal lattice, leading to larger onsite energies.
In both layers, the in-plane crystal-field splitting parameters are well approximated using three Gaussian functions as
\begin{equation}\label{Gaussian}
    V(\mathbf{r}_\perp) = V_0[2 e^{-(\mathbf{r}_\perp-\mathbf{\mu}_\mathrm{AA})^2/\sigma^2} - e^{-(\mathbf{r}_\perp-\mathbf{\mu}_\mathrm{AB})^2/\sigma^2} - e^{-(\mathbf{r}_\perp-\mathbf{\mu}_\mathrm{BA})^2/\sigma^2}],
\end{equation}
where $\mathbf{r}_\perp$ is the in-plane component of the radius vector, $\mathbf{\mu}_\mathrm{AA}=(\mathbf{t}_1 + \mathbf{t}_2)/3$ is the center of the AA-stacked region, and $\mathbf{\mu}_\mathrm{AB}=0$ and $\mathbf{\mu}_\mathrm{AB}=2(\mathbf{t}_1 + \mathbf{t}_2)/3$ are the centers of the AB- and BA-stacked regions, respectively. This expression has to be summed over the adjacent unit cells to take into account their contributions. 

Unfortunately, the first-principles methodology employed in our study is computationally too expensive to be applied to magic-angle TBG. Our results can nevertheless be extended to small twist angles when strong lattice relaxation result in AB and BA domains covering most of the moiré supercell \cite{gargiulo2017structural}. Within such domains the crystal-field parameters are those of properly stacked bilayer graphene, and are expected to change smoothly across the stacking domain boundaries.

\begin{figure*}[t!]
	\centering
	\includegraphics[width=0.85\textwidth]{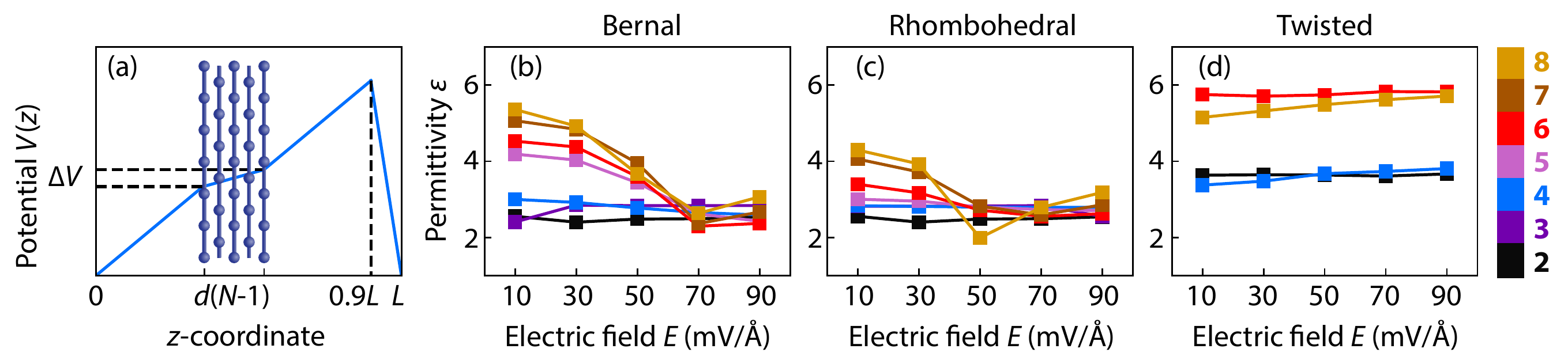}
	\caption{(a) Behaviour of electric potential $V(z)$ along the $z$-direction within the supercell used in DFT calculations and dielectric permittivity of (b) Bernal-stacked, (c) rhombohedral-stacked and (d) twisted multilayer graphene as a function of electric field strength. Different colors in (b)--(d) encode the number of layers in the system.}
	\label{screening}
\end{figure*}

In many-layer systems with $N\geq4$, the crystal-field pattern at the twist boundary becomes more complicated. In addition to the moir\'{e}-modulated splitting described by Eq.~(\ref{Gaussian}), the onsite energy of each atom depends on whether there is an atom with the same lateral position in the adjacent layer on the other side. On the other layers (not shown in the figure), the crystal field is practically identical to that in the Bernal-stacked structures shown in Fig.~\ref{crystalOrdered}(b), so that there a lower onsite energy on those atoms which are stacked on top of each other. The moir\'{e}-modulated crystal field thus appears almost exclusively on the pair of mutually rotated layers, although the twist affects the crystal field values in the entire structure.

It is still possible to describe the in-plane crystal field in tMLG by employing the following approach. We fit Bernal-like (staggered) splitting in each layer $n$ by setting negative potentials $-\delta_n$ on those atoms which have an atom on top in the adjacent layer, $\delta_n$ otherwise. In case of the middle two layers, we simultaneously fit $\delta_n$ and parameters $V$ and $\sigma$ from Eq.~(\ref{Gaussian}). The fitting results for all the twisted multilayer are presented in Table~\ref{Fitting}.
We note that the Bernal-like splitting in the four- and six-layer tMLG is very similar. It grows with the period of the moir\'{e} lattice and increases towards the inner layers of the structure. On the other hand, the in-plane splitting in the eight-layer system has an opposite sign to that of the other twisted structure and all Bernal-stacked ordered multilayers. The width $\sigma$ of the Gaussian-like splitting naturally increases with the period of the superlattice, and the amplitude $V$ follows approximately the same trend. 

\begin{table*}[t]
    \caption{In-plane crystal-field splitting parameters in twisted multilayer graphene. Here, $N$ is the number of layers, $i$ is the rotation index, $\delta_n$ are the Bernal-like splitting parameters in the $n$-th layer starting from the outside,
    $V_0$ and $\sigma$ are the fitting parameters for the Gaussian in-plane splitting on the twist boundary, and MAE is the mean absolute error of the fitting. The Bernal-like splitting parameters are only given for half of the layers, while in the other half the values are the same for symmetry reasons.}
    \centering
    \begin{tabular}{@{}rrrrrrrr@{}}
        \hline
\hline
        Structure & $V_0$ (meV) & $\sigma$ (\AA) & $\delta_1$ (meV) & $\delta_2$ (meV) & $\delta_3$ (meV) & $\delta_4$ (meV) & MAE (meV) \\
\hline
        \multicolumn{1}{l}{$N=2$} \\
        $i=1$ & 4.4 & 1.4 & -- & -- & -- & -- & 0.1  \\
        $i=2$ & 9.3 & 4.6 & -- & -- & -- & -- & 0.2  \\
        $i=3$ & 7.0 & 5.1 & -- & -- & -- & -- & 0.1  \\
        \multicolumn{1}{l}{$N=4$} \\
        $i=1$ & 8.9 & 2.9 & 0.9 & 1.1 & -- & -- & 0.3  \\
        $i=2$ & 8.5 & 4.6 & 2.0 & 1.4 & -- & -- & 1.4 \\
        \multicolumn{1}{l}{$N=6$} \\
        $i=1$ & 7.8 & 2.9 & 0.9 & 2.2 & 1.3 & -- & 0.1 \\
        $i=2$ & 10.1 & 4.7 & 1.5 & 3.0 & 2.0 & -- & 0.5 \\
        \multicolumn{1}{l}{$N=8$} \\
        $i=1$ & 6.6 & 1.1 & $-1.4$ & $-0.8$ & $-1.2$ & 1.1 & 0.6 \\
\hline
\hline
        \end{tabular}
    \label{Fitting}
\end{table*}

It has to be noted that the above crystal fields were calculated for the configurations suspended in vacuum. Encapsulation in $h$-BN commonly employed in experiments is expected to provide additional contributions in the form of crystal fields. \cite{pasquier2019crystal}. Another factor that could modify the crystal fields in multilayer graphene is the local charge carrier concentration, for example, modulated by the patterned gate voltage. The impact of both factors on the crystal field effect in MLGs will be investigated in our future studies.

\textbf{Electric-Field Screening.} Next we use the same \textit{ab initio} approach to study the screening of external electric field in multilayer graphene. Out-of-plane electric field is included in the DFT calculations as a sawtooth potential across the simulation cell periodically repeated along the out-of-plane direction as depicted in Figure~\ref{screening}(a). The slope of this potential naturally reduces inside the multilayer due to dielectric screening. The results of the DFT calculations are then processed using Wannier90 the same way as before. Note that the onsite potentials obtained during this procedure include both the induced electric polarization and internal crystal-field splitting. It is thus crucial to adjust the onsite potentials by subtracting the crystal-field splitting parameters, isolating the polarization coming solely from external field.

The dielectric permittivity of an $N$-layer system is then defined as
\begin{equation}\label{epsilon}
    \varepsilon = \frac{eEd(N-1)}{\Delta V},
\end{equation}
where $d(N-1)$ is the multilayer thickness and $\Delta V$ is the difference between average potentials on the first and last layers. Note that the value of $\varepsilon$ depends not only on the electric field strength and on MLG thickness, but also on the stacking order of graphene sheets.  Figures~\ref{screening}(b) and (c) compare the dielectric permittivity of ABA- and ABC-stacked multilayers for electric-field strengths in the range 10--90 mV/\AA.

One can see that in the weak-field limit, the dielectric permittivity increases with the thickness of multilayer graphene. This is consistent with the fact that thicker multilayers have larger out-of-plane dipole moments and thus higher polarizability in the $z$-direction \cite{tepliakov2016field}. In case of the structures with $N\leq4$ layers, the dielectric response is practically identical for the Bernal- and rhombohedral-stacked multilayers. There, $\varepsilon$ is mostly independent of the electric field strength and has values around 2.5--3.0. On the contrary, in configurations with $N\geq5$ layers, $\varepsilon$ is highly sensitive to the strength of external electric field and is also notably larger in the case of Bernal stacking. 
In general, for both types of stacking dielectric permittivity reduces and then saturates upon increasing the strength of electric field $E$. 
As illustrated with an effective self-consistent model in the Supplementary information, this behavior is due to the saturating electronic polarization in strong electric fields (see Figure~S1). MLG with $N\geq6$ layers also features a pronounced minimum in the $\varepsilon$ plots. For example, dielectric permittivity of eight-layer ABC-stacked graphene reaches its minimal value of 2.0 for an electric field strength of 50~mV/\AA. From the shape of the plots, it can be expected that similar minima could be observed for the structures with smaller number of layers, but for larger electric field values, which are inaccessible in the computational approach employed here. The nature of this minimum in many-layer structure is unclear, but it could be attributed to the enhancement of electron-electron interaction at large induced densities.

The electric field screening in tMLG is quantified in the same way as in the case of ordered MLG. Since the onsite potentials at the twist boundary of these structures are strongly modulated by the moir\'{e} superlattice (see Figure~\ref{crystalTwisted}), it would be natural to expect that the dielectric response of tMLG varies in the $xy$-plane of the multilayer. However, our calculations show that these variations are very weak, and the electric field screening in twisted multilayers can also be characterized by a single parameter $\varepsilon$ defined in Eq.~(\ref{epsilon}), just as in ordered MLG models.
Figure~\ref{screening}(d) shows the dielectric permittivity of the studied tMLG configurations as a function of electric field strength. We only considered structures with the first twist angle $\theta=21.8^\circ$, because the addition of external electric field to the first-principles calculation increases their complexity and smaller twist angles would demand unreasonable computation costs. According to the figure, the variations of $\varepsilon$ in tMLG are much less pronounced than in case of ordered multilayer configurations. It is thus possible to introduce twist into thick MLG to eliminate the dependence of its dielectric permittivity on the field strength, if this is desirable for the electronic applications.

Furthermore, the overall values of dielectric permittivity are also larger than in case of ordered systems, reaching 4 in case of twisted bilayer and double bilayer graphene, and 6 in case of six- and eight-layer structures. Larger values of $\varepsilon$ indicate higher polarisability of twisted multilayers, which may be attributed to the redistribution of electrons in the moir\'{e} lattice upon the application of external electric field. The dielectric permittivity values for all analyzed structures are provided in Table~S6 of the Supplementary Information.


To conclude, we have comprehensively analyzed the crystal field effect and electric field screening in a broad range of ordered and twisted configurations of multilayer graphene. By using an approach combining \textit{ab initio} calculations with the Wannier function formalism, we obtain the onsite potentials of carbon atoms in the investigated MLG models. It was shown that in the absence of external electric field, MLG features intrinsic electronic polarization due to the crystal field effect. In ordered MLG consisting of ABA- and ABC-stacked sequences the crystal field is mostly directed towards the exterior layers of the structure, and the interlayer splitting has values $\sim20$~meV. The in-plane component of the crystal field is much less pronounced in Bernal-stacked MLG ($\sim6$~meV) and practically negligible in rhombohedral-stacked MLG. In contrast, twisted MLG configurations feature a strong in-plane crystal field $\sim12$~meV modulated by the moir\'{e} potential. The out-of-plane intrinsic field is weaker in tMLG ($\sim10$~meV) and directed towards the center of the multilayer. The dielectric permittivity of the ABA- and ABC-stacked MLG is strongly nonlinear with respect to the electric field strength, varying between $2.0$ and $5.4$, whereas the dielectric permittivity of tMLG is practically independent of $E$, having values between $3.4$ and $5.8$ for different structures. The calculated parameters are rationalized using simple models and tabulated in the Supplementary material. 
The findings of our study thus provide significant theoretical insights into the physics of graphene multilayers and will be of benefit for the design of new electronic devices based on multilayer graphene.

\vskip 1cm

{\bf Methods}
Electronic properties of graphene multilayers were calculated using density functional theory (DFT) with the plane-wave basis set as implemented in Quantum ESPRESSO \cite{giannozzi2009quantum}. Our DFT calculations were carried out using energy and charge cutoffs of 45~Ry and 360~Ry, respectively, 15$\times$15$\times$1 k-grid, and a simulation cell dimension of 50~\AA\ in the $z$-direction. In the case of twisted structures with $N=4$, $i=3$ and $N=8$, $i=2$, the k-grid was reduced to 9$\times$9$\times$1 to decrease the computational complexity.

External electric field was introduced into DFT calculations as a sawtooth potential varying in the $z$-direction. This potential was made to increase along 9/10 of the supercell and decrease to the initial value in the rest of the supercell, in the middle of the vacuum region, in order to keep the overall potential periodic along the $z$-direction (see Figure~\ref{screening}(a)). 
The region where the external potential decreases was placed in such a way as to prevent the electron density from accumulation in the potential well that is formed by the sawtooth potential in the vacuum between the layers.

The Bloch states obtained from the DFT calculations were then processed with Wannier90 \cite{mostofi2008wannier90} to construct the projector Wannier functions using the $p_z$ orbitals of C atoms. This procedure yields an \textit{ab initio} tight-binding Hamiltonian, the diagonal elements of which correspond to the onsite energies of the $p_z$ orbitals. These energies are then used to analyze the crystal-field effect in multilayer graphene, as well as the screening of external electric field, as described in the main text.

{\bf Supporting Information}
Effective analytical models providing qualitative description of the electric field screening and intrinsic symmetric polarization in graphene multilayers. Tabulated data for Figures~\ref{crystalOrdered}--\ref{screening} of the main text. 

{\bf Acknowledgments}
This research was supported by the NCCR Marvel, funded by the Swiss National Science Foundation (project No. 51NF40-182892). N.V.T. acknowledges the Scholarship of the President of the Russian Federation for Abroad Education. Calculations were performed at the facilities of Scientific IT and Application Support Center of EPFL and the Swiss National Supercomputing Centre (CSCS) under project s1008.


%

\end{document}